\documentclass[referee]{aa}
\usepackage{graphicx}
\begin{document}


  \title{Collisional depolarization and transfer rates of spectral lines by atomic hydrogen. II: application to $d$ states of neutral atoms.}

   \author{M. Derouich
          \inst{1},
           S. Sahal-Bréchot
          \inst{1},
             and
          P. S. Barklem
          \inst{2}
          }
\titlerunning{Collisional depolarization  and transfer rates.}
\authorrunning{M. Derouich et al}
   \institute{Observatoire de Paris-Meudon, \textbf{LERMA UMR CNRS 8112}, 5, Place Jules Janssen, F-92195 Meudon Cedex, France. 
          \and
Department of Astronomy and Space Physics, Uppsala University, Box 515, S 
751 20 Uppsala, Sweden\\
   \email{Moncef.Derouich@obspm.fr}
             }
   \date{Received 2003 / accepted 2003}

\abstract{
The theory of collisional depolarization of spectral lines by atomic hydrogen
(Derouich et al.  \cite{derouich1}) is extended to $d$ $(l$=2) atomic levels. Depolarization rates, polarization and population transfer rates are calculated and results are given as a function of the temperature. Cross sections as a function of the effective quantum 
number for a relative velocity of $10$ $\textrm{km s}^{-1}$  are also given together with velocity exponents $\lambda$, if \textbf{they exist}, on the assumption that the cross section varies with velocity as $v^{-\lambda}$. A discussion of our results is presented. 

}
\maketitle


\section{Introduction}
In  stellar atmospheres,  isotropic collisions with the particles of the medium 
are responsible for a part of  the broadening of  spectral lines.  If a polarizing 
effect creates an atomic polarization of the lines, they can  decrease or even destroy this atomic
polarization. Spectral lines observed at the solar limb are linearly polarized
by anisotropic scattering of the incident solar radiation. Isotropic collisions with the  particles of the medium depolarize the lines. In the solar photosphere and the low chromosphere,  depolarizing collisions are dominated by isotropic collisions with hydrogen atoms. When these depolarizing collisions dominate over any other radiative (or collisional) polarizing effects, the 
atomic levels become depolarized. 
Therefore  \textbf{depolarization rates, polarization and population transfer rates} by collisions with hydrogen are needed in order to interpret the observed polarization.

In Derouich et al. (2003), referred to hereafter as Paper I, a semi-classical theory for depolarization of neutral atomic lines by collisions with atomic hydrogen has been developed and applied to $p$ ($l=1$) atomic states. \textbf{This} theory is an extension to collisional depolarization of the theory developed in the  1990's by O'Mara and collaborators (Anstee \cite{Anstee2}; Anstee \& O'Mara \cite{Anstee1}, \cite{Anstee3}; Anstee, O'Mara \& Ross \cite{Anstee4}; Barklem \cite{Barklem3}; Barklem \& O'Mara \cite{Barklem1}; Barklem, O'Mara \& Ross \cite{Barklem2}) for line broadening by collisions with atomic hydrogen. The  present paper is an  extension of \textbf{this} theory to $d$ ($l=2$) atomic levels. In fact, this paper presents the first calculations of the depolarization  and the collisional transfer rates for $d$-atomic states.

A great advantage of the present method is that calculations are not specific for a given perturbed atom. The transition  matrix $T$  may be calculated using Coulomb wavefunctions  for the valence electron of the perturbed atom \textbf{and is} simply dependent on the effective principal number $n^*$ and the orbital angular momentum quantum number $l$ ($l$=2 for $d$-atomic states).  Therefore we can calculate  \textbf{depolarization cross sections,  polarization and population transfer cross sections} for any level of any atom, allowing computation for complex atoms. This
is very useful for interpreting the so-called ``second solar spectrum'' (Stenfo \& Keller \cite{Stenflo}), where depolarization rates for many levels are needed
(Manso Sainzo \& Landi Degl'Innocenti \cite{manso}).

An extension of our theory to higher $l$-values, \textbf{aimed at}  a more complete 
interpretation of the second solar spectrum, will be the subject of further 
papers.

\section{Description of the problem and summary of Paper I}
In our collision problem, the perturbed atoms collide with a bath of perturbing
hydrogen atoms. The hydrogen atom is assumed to remain in its ground state during the 
collision. The internal states of the perturbed atom are  described by the 
spherical tensor components $\displaystyle ^{nl J}\rho_q^k$ of 
the density matrix. Due to the isotropy of the collisions, \textbf{the depolarization rates,  polarization and population transfer rates} are $q$-independent. The contribution of the isotropic collisions to the statistical 
equilibrium equations is: 
\begin{eqnarray} \label{eq1}
\big(\frac{d \; ^{n l J}\rho_0^{k}}{dt})_{coll} & = & - D^k(n l J, T) \; ^{n l J}\rho_0^k 
+ \sum_{J \ne J'}  
D^k(n l J \to n l J', T) \;  ^{n l J'}\rho_0^k  \\  
&+&\textrm{quenching term,} \nonumber
\end{eqnarray}  
where  $\displaystyle D^k(n  l J, T)$  is the collisional 
depolarization  rate for the statistical tensor of rank $k$. Each 
level of total angular momentum $J$ relaxes with $2J+1$ independent 
depolarization  rates. In particular $D^0(nl J, T)$ is the destruction rate of
 population, $D^1(nl J, T)$ is the destruction rate of orientation 
(circular atomic polarization) and $D^2(nl  J, T)$ is the destruction rate of 
alignment which is of interest in the understanding of the second solar 
spectrum. 

We assume that  inelastic collisions with hydrogen perturbers do not alter the total population of an atomic level ($\displaystyle {n l J}$). This is the so-called no-quenching approximation. The no-quenching approximation implies that 
\begin{eqnarray} \label{eq2}
D^0(n l J, T)= \sum_{J \ne J'}\zeta (nl J  \to nl J', T) =0,
\end{eqnarray}
 where  $\zeta (nl J \to nl J', T) (J \ne J')$ is the inelastic collisional rate.
The expression for the depolarization rate $D^k(nl  J, T)$ is given in Paper I.

If the quenching must be taken into account, $\displaystyle D^k(n  l J \to n  l J', T)$ corresponds to 
collisional transfer of population $(k=0)$, orientation $(k=1)$ and alignment $(k=2)$.   $D^k(n l J \to n l J', T)$ can be written as a 
linear combination of the collisional transition rates between the fine stucture sublevels $\zeta (n l J M_J \to n l J' M'_J, T) (J \ne J')$ (Sahal-Bréchot \cite{Sahal2}): 
\begin{eqnarray} \label{eq3}
D^k(n l J \to n l J', T) =(2k+1)&& \sum_{M_J,M'_J} (-1)^{J+J'-M_J-M'_J}
  \left(
\begin{array}{ccc} 
J & k &  J  \\
-M_J &  0 & M_J  
\end{array}
\right) \\
&&\times \left(
\begin{array}{ccc} 
J' & k &  J'  \\
-M'_J &  0 & M'_J  
\end{array}
\right) \zeta (nl J M_J \to nl J' M'_J, T). \nonumber
\end{eqnarray}
The expressions between parentheses denote $3j$-coefficients (Messiah \cite{Messiah}). 

In particular one obtains
\begin{eqnarray} \label{eq4}
D^0(n l J \to n l J', T) = \sqrt{\frac{2J+1}{2J'+1}} \zeta (n l J \to n l J', T), 
\end{eqnarray}
where 
\begin{eqnarray} \label{eq5}
\zeta (n l J \to n l J', T)& = & \frac{1}{2J+1}  
 \sum_{M_J,M_J'} \zeta (n l J M_J \to n l J' M'_J, T), 
\end{eqnarray}
is the collisional transition rate between fine structure levels. $\zeta (\alpha J  M_J \to \alpha J  M'_J, T)$ is the collisional transition rate between the sublevels $|\alpha J M_J \rangle \to |\alpha J  M'_J \rangle$. It 
can be written as a function of the local temperature $T$ and the hydrogen 
perturber local density $n_H$ (Paper I):
\begin{eqnarray} \label{eq6}
\zeta(\alpha J M_J \to \alpha J M'_J, T) = n_H \int_{0}^{\infty} \int_{0}^{\infty} 2\pi b\, db\: v\; f(v)\; dv\:  |\langle \alpha J M_J| I - S(b, v) |\alpha J M'_J \rangle|^2,
\end{eqnarray}  
where $I$ is the unit matrix and $T = I-S $ is the so-called 
transition matrix depending on the impact-parameter $b$ and relative velocity $v$. \textbf{Then} the collisional depolarization rates and the 
collisional transfer rates can 
be expressed in terms of the $S$-matrix elements for the collision 
which are functionally dependent on the interaction energy of hydrogen in its 
ground state  with the perturbed atom. The essential difference between 
various theoretical  computations of the depolarization 
and  collisional transfer rates  is in the method employed to determine the 
interaction energy and to determine the $S$-matrix.  
\section{Method}
Consider a collision between a perturbed atom in a $l=2$ state and hydrogen 
in its ground state $1s$.  The perturbed atom is described by an optical 
electron outside a positively charged core. The effective principal number
is given by $ n^*=[2 (E_\infty - E_{n l })]^{-1/2}$. $\displaystyle E_{n l }$ is the energy of the state of the valence electron and $\displaystyle E_\infty$ is  the appropriate series limit for the parent configuration of the perturbed atom state.  As  in the $p$ atomic levels case (Paper I), depolarizing 
collisions are due essentially to intermediate-range isotropic interactions 
between radiating and hydrogen atoms. The so-called Rayleigh-Schrödinger-Unsöld (RSU) interaction potential, used in this work, is of semi-classical nature and is 
totally electrostatic. For more details we refer to Paper I and to the ABO 
papers (Anstee \cite{Anstee2}; Anstee \& O'Mara \cite{Anstee1}, \cite{Anstee3}; Anstee, O'Mara \& Ross \cite{Anstee4}; Barklem \cite{Barklem3}; Barklem \& 
O'Mara \cite{Barklem1}; Barklem, O'Mara \& Ross \cite{Barklem2}). We  use
the \textbf{notation} of Paper I in the following \textbf{analysis}.

Inelastic collisions which leave the radiating atom in a final state 
$(n'l')$  different from the initial one $(nl)$ are neglected. Thus we consider only the \textbf{$(2l+1)$ states of the subspace $(nl)$}. Since  potentials are computed in the rotating frame (Paper I), which is obtained from  the fixed laboratory frame by means of the geometrical rotation $R$ $(\beta, \frac{\pi}{2},\frac{\pi}{2})$, the interaction potential is 
diagonal. The $(2l+1)$ RSU potential elements calculated here are (Anstee \& O'Mara  \cite{Anstee2}) 
\begin{eqnarray} \label{eq7}  
 V_{eff,m_l}& =& _{\scriptstyle(H)}\langle 1 0 0| _{\scriptstyle (A)}\langle n l m_l  | V_{eff}  |n l m_l \rangle_{\scriptstyle (A)}|1 0 0 \rangle_{\scriptstyle(H)} \\
& =& \langle M_l \big\vert V \big\vert M_l \rangle   - \frac{1}{E_p}(\langle M_l \big\vert V\big\vert M_l \rangle )^2 
+ \displaystyle \frac{1}{E_p} \int_{0}^{+\infty} P^2_{n^*l} I_{l|m_l|}(R,p_2) 
dp_2, \nonumber
\end{eqnarray}  
in atomic units which are used hereafter. Here $\big\vert M_l \rangle = |n l m_l \rangle_{\scriptstyle (A)}|1 0 0 \rangle_{\scriptstyle(H)}$,   $\displaystyle I_{l|m_l|}$ are lengthy complicated analytic 
functions and $E_p = - 4/9$ is adopted. $\displaystyle P_{n^*l}$ is the Coulomb radial wave function for the valence electron of 
the perturbed atom  with  quantum defect $\displaystyle \delta = n - n^*$. 

The total wavefunction  
$|\psi \rangle$ of the system (atom+perturber) is taken as the product of the wave function $|\psi \rangle_{\scriptstyle (A)}$ of the perturbed atom and 
 that of hydrogen in its ground state $|1 0 0 \rangle_{\scriptstyle (H)}$:
\begin{eqnarray} \label{eq8}
|\psi \rangle =|\psi \rangle_{\scriptstyle (A)}|1 0 0 \rangle_{\scriptstyle (H)},
\end{eqnarray} 
\textbf{and} is developed over the basis formed by the products of the separated atoms states $ | M_l \rangle $:
\begin{eqnarray} \label{eq9}
\big | \psi (t) \rangle = \displaystyle \sum_{M_l} a_{M_l}(t)   \textrm{e}^{-\textrm{i}{E^0_{M_l} t}} \big | M_l \rangle 
\end{eqnarray}
\textbf{where} $\displaystyle E^0_{M_l}$ is the eigenenergy of the system made-up of the two 
isolated atoms. 

The semi-classical coupled linear differential equations are obtained \textbf{from} 
the time-dependent Schr\"odinger equation.   The transformation from the rotating to the fixed laboratory frame is included following Roueff (\cite{roueff}).  For $d$-states,  the coupled differential 
equations become explicitly (Barklem \& O'Mara  \cite{Barklem1}):
\begin{eqnarray} \label{eq10}
\textrm{i} \frac{\partial a_{-2}(t)}{\partial t} &= &\frac{1}{8} a_{-2}(t) (3 V_{eff,0} + 4 V_{eff,1} + V_{eff,2})+ \frac{3}{4 \sqrt6} a_{0}(t) \textrm{e}^{- 2\textrm{i}\beta} (-V_{eff,0} + V_{eff,2}) \nonumber \\& + &\frac{1}{8} a_{2}(t) \textrm{e}^{- 4\textrm{i}\beta} (3 V_{eff,0} - 4 V_{eff,1} + V_{eff,2}) \nonumber \\
\textrm{i} \frac{\partial a_{-1}(t)}{\partial t} & = & \frac{1}{2} a_{-1}(t) (V_{eff,1} + V_{eff,2}) +\frac{1}{2} a_{1}(t) \textrm{e}^{- 2\textrm{i}\beta} (-V_{eff,1} + V_{eff,2}) \nonumber \\ 
\textrm{i} \frac{\partial a_{0}(t)}{\partial t} &= &\frac{3}{4 \sqrt6} a_{-2}(t) \textrm{e}^{2\textrm{i}\beta} (-V_{eff,0} + V_{eff,2})
+ \frac{3}{4 \sqrt6} a_{2}(t) \textrm{e}^{- 2\textrm{i}\beta} (-V_{eff,0} + V_{eff,2}) \ \\ &+ & \frac{1}{4} a_{0}(t) ( V_{eff,0} + 3 V_{eff,2}) \nonumber \\
\textrm{i} \frac{\partial a_{1}(t)}{\partial t}&=&\frac{1}{2} a_{-1}(t) \textrm{e}^{ 2
\textrm{i}\beta} (- V_{eff,1} + V_{eff,2})+\frac{1}{2} a_{1}(t) (V_{eff,1} + 
V_{eff,2}) \nonumber \\
\textrm{i} \frac{\partial a_{2}(t)}{\partial t} &=& \frac{1}{8} a_{-2}(t) \textrm{e}^{4\textrm{i}\beta} (3 V_{eff,0} - 4 V_{eff,1} + V_{eff,2})+ \frac{3}{4 \sqrt6} a_{0}(t) \textrm{e}^{ 2\textrm{i}\beta} (-V_{eff,0} + V_{eff,2}) \nonumber \\& +& \frac{1}{8} a_{2}(t) (3 V_{eff,0} + 4 V_{eff,1} + V_{eff,2}) \nonumber 
\end{eqnarray}
Having the interaction potential $\displaystyle V_{eff}$, after integration of these equations,  we obtain 
the transition matrix elements  in the $| n l  m_l \rangle$ basis 
for a given velocity and  impact parameter. The $T$-matrix elements in the $| n l J M_J \rangle$ basis, which are needed for the 
depolarization and collisional transfer rates calculations, are obtained from  equation (21) of Paper I. 

In the \textbf{irreducible} tensorial \textbf{operator} basis, the angular average over all 
possible directions of the collision plane of  the depolarization transition  probability is given in Paper I. That of the collisional transfer
 transition probability is given by:
\begin{eqnarray} \label{eq11}
\langle P^k(n l J \to n l J', b, v)  \rangle_{av}& = &
\sum_{\mu,\mu' , \nu,\nu'} \langle n l \; J \; \mu |T|n l \; J' \; \mu'\rangle  \langle n l \; J \; \nu |T|n l \; J' \; \nu'\rangle^* \\
&&\times \quad \sum_{\chi} (-1)^{J-J'+\mu-\mu'}
\left(
\begin{array}{ccc} 
J & J &  k  \\
\nu &  -\mu & \chi  
\end{array}
\right)
\left(
\begin{array}{ccc} 
J' & J' &  k \\ 
\nu' &  -\mu' & \chi  
\end{array}
\right). \nonumber 
\end{eqnarray}
Owing to the selection rules for the $3j$-coefficients, the summation over $\chi$ is reduced to a single term, since $\chi = -(\nu'-\mu')=-(\nu-\mu)$. 

The depolarization rates \textbf{$\displaystyle D^k(n l  J, T)$, and the polarization and  population transfer rates} $\displaystyle D^k(n l  J \to n l J', T)$   follow from  integration over the 
impact parameters and the velocities with a Maxwellian distribution (for 
more details see Paper I). 

\section{Results}
In  most cases, the behaviour of the cross sections with the relative velocity $v$ obeys a power law of the form: 
\begin{eqnarray} \label{eq12}
\sigma^k(n l J \to n l J', v) (J=J'\; \textrm{and}\; J\ne J')= 
\displaystyle \sigma^k(n l J \to n l J', v_0)(\frac{v}{v_0})^{-\lambda^k(n l J \to n l J')} ,
\end{eqnarray}
where ${v_0}$ is a typical velocity where the cross section is calculated ($10$ km s$^{-1}$). In certain cases  here, such behaviour was not obeyed (the cross section showed oscillations with velocity). 

Table \ref{table1}  gives the various cross sections as function of $n^*$ for a relative velocity of $10 \;\textrm{km} \; \textrm{s}^{-1}$ and 
the corresponding velocity exponents $\displaystyle \lambda^k(n l J \to n l J')$,  if the exponential \textbf{behaviour} was obeyed,  are tabulated in  Table \ref{table2}.  Then we can obtain the cross sections $\displaystyle \sigma^k(n l J \to n l J', v)$ for all velocities from Tables 1 and 2 using equation (\ref{eq12}).  Tables \ref{table1} and \ref{table2}   can be interpolated for an appropriate $n^*$  associated to a given observed line  in order to obtain the needed  \textbf{depolarization  cross sections  and  collisional transfer cross-sections. After integration over velocities of these cross-sections, we obtain the  depolarization rates and the collisional transfer  rates} of the  line studied. 

For cross sections obeying  equation (\ref{eq12}), as in Anstee \& O'Mara (\cite{Anstee2}),  the depolarization and the collisional transfer rates   
 can be expressed by: 
\begin{eqnarray} \label{eq13}
\displaystyle D^k(n l J \to n l J', T) (J=J' \; \textrm{and} \; J\ne J')& =&(\frac{4}{\pi})^{(\frac{1}{2}\lambda^k{(n l J \to n l J')})} 
 \Gamma (2-\frac{1}{2} \lambda^k(n l J \to n l J')) \nonumber \\
& \times &   v_0   \sigma^k(n l J \to n l J', v_0) 
 (\frac{\bar{v}}{v_0})^{1-\lambda^k(n l J \to n l J')} 
\end{eqnarray} 
We can generalize this relationship, which is specific to a particular atom owing to
its mass, by assuming that $\mu=m_H$ ($\mu$ and $m_H$ are the reduced and 
hydrogen mass respectively). This approximation introduces a very small error 
(Paper I).

\begin{table}
\begin{center}
\begin{tabular}{|l|c|c|c|c|c|r|}
\hline
$n^*$ & $\sigma^2(n 2 \frac{3}{2})$ &$\sigma^2(n 2 2)$& $\sigma^2(n 2 \frac{5}{2})$ & $\sigma^0(n 2 \frac{3}{2} \to n 2 \frac{5}{2})$ & $\sigma^2 (n 2 \frac{3}{2} \to n 2 \frac{5}{2})$\\
\hline
$2.5$ &283 & 507& 342& 278&60	 \\
\hline 
$2.6$ &313& 566& 380&311 &65  	\\
\hline 
$2.7$ &	351& 633& 424&351 &73	\\
\hline 
$2.8$&394	& 717& 477&397 &77	\\
\hline
$2.9$&	443& 808& 537&451 & 87	\\
\hline 
3&496& 900	&598 &512 &104\\
\hline
$3.1$&553&1008	 &666	&578 &116 \\
\hline 
$3.2$ &628 &1109 &727	&647 & 139 \\
\hline 
$3.3$ &711  & 1210&800	&711 & 163\\
\hline 
$3.4$ &812& 1348&897	&783 &179 \\
\hline 
$3.5$ & 905&1470	&964	&847 &190 \\
\hline 
$3.6$ & 1028&1621	&1039	&920 &196\\
\hline 
$3.7$ &	1108&1763& 1106	&961 &183\\
\hline 
$3.8$ &1202& 1910&1185	&1004 &195\\
\hline 
$3.9$ &1299&2066 &1260	&1079 &206\\
\hline 
$4$ &   1294&2191	&1367	&1117 &232\\ 
\hline  
\end{tabular}
\end{center}
\caption{Variation of the  cross sections, for the relative velocity  of $10 \;\textrm{km} \; \textrm{s}^{-1}$,  with the effective principal number. Cross sections are in atomic units.}
\label{table1}
\end{table} 

\begin{table}
\begin{center}
\begin{tabular}{|l|c|c|c|c|r|}
\hline
$n^*$ & $\lambda^2(n 2 \frac{3}{2})$ & $\lambda^2(n 2 2)$& $\lambda^2(n 2 \frac{5}{2})$ & $\lambda^0(n 2 \frac{3}{2} \to n 2 \frac{5}{2})$ & $\lambda^2 (n 2 \frac{3}{2} \to n 2 \frac{5}{2})$\\
\hline
$2.5$ &0.207 & 0.258& 0.268 &0.259	&0.172	\\
\hline 
$2.6$ &0.223 & 0.260& 0.269 &0.258  & 0.184 	\\
\hline 
$2.7$ &0.222	&0.253& 0.268 &0.259& 0.233	\\
\hline 
$2.8$&0.200&0.253& 0.271 &0.257	& 0.209	\\
\hline
$2.9$&0.214& 0.261& 0.270&0.253	&0.177	\\
\hline 
3&0.210 &0.265&0.280	&0.249	&0.148 \\
\hline
$3.1$&0.185&0.273&0.282 &0.245& - \\
\hline 
$3.2$ &0.168&0.274&0.284	&0.257&- \\
\hline 
$3.3$ &0.153&0.255	&0.280	&0.267&0.194\\
\hline 
$3.4$ &0.137&0.253	&0.285	&0.279&0.238 \\
\hline 
$3.5$ &0.140&0.258	&0.304	&0.286&0.306 \\
\hline 
$3.6$ & 0.131&0.247	&0.315	&0.272& 0.394\\
\hline 
$3.7$ &0.175&0.257	&0.330 &0.249&0.446\\
\hline 
$3.8$ &	0.179&0.250 &0.303	&0.234&0.479\\
\hline 
$3.9$ &	0.208&0.283 &0.322 &0.237&0.399\\
\hline 
$4$ &0.172&0.262	&0.296	&0.226&0.448\\ 
\hline  
\end{tabular}
\end{center}
\caption{Velocity exponents $\displaystyle \lambda^k(n l J \to n l J') (J=J' \; \textrm{and} \; J\ne J')$ corresponding to the cross sections of Table \ref{table1}.}
\label{table2}
\end{table}

As mentioned, in certain cases especially for transfer of linear polarization calculations, the cross sections do not show exponential behaviour with velocity and so $\lambda$ is not reported (Table \ref{table2}). In these cases the linear polarization transfer rates must be computed by a numerical integration over computed cross sections obtained at different velocities.   

\begin{figure}[htbp]
\begin{center}
\includegraphics[width=8 cm]{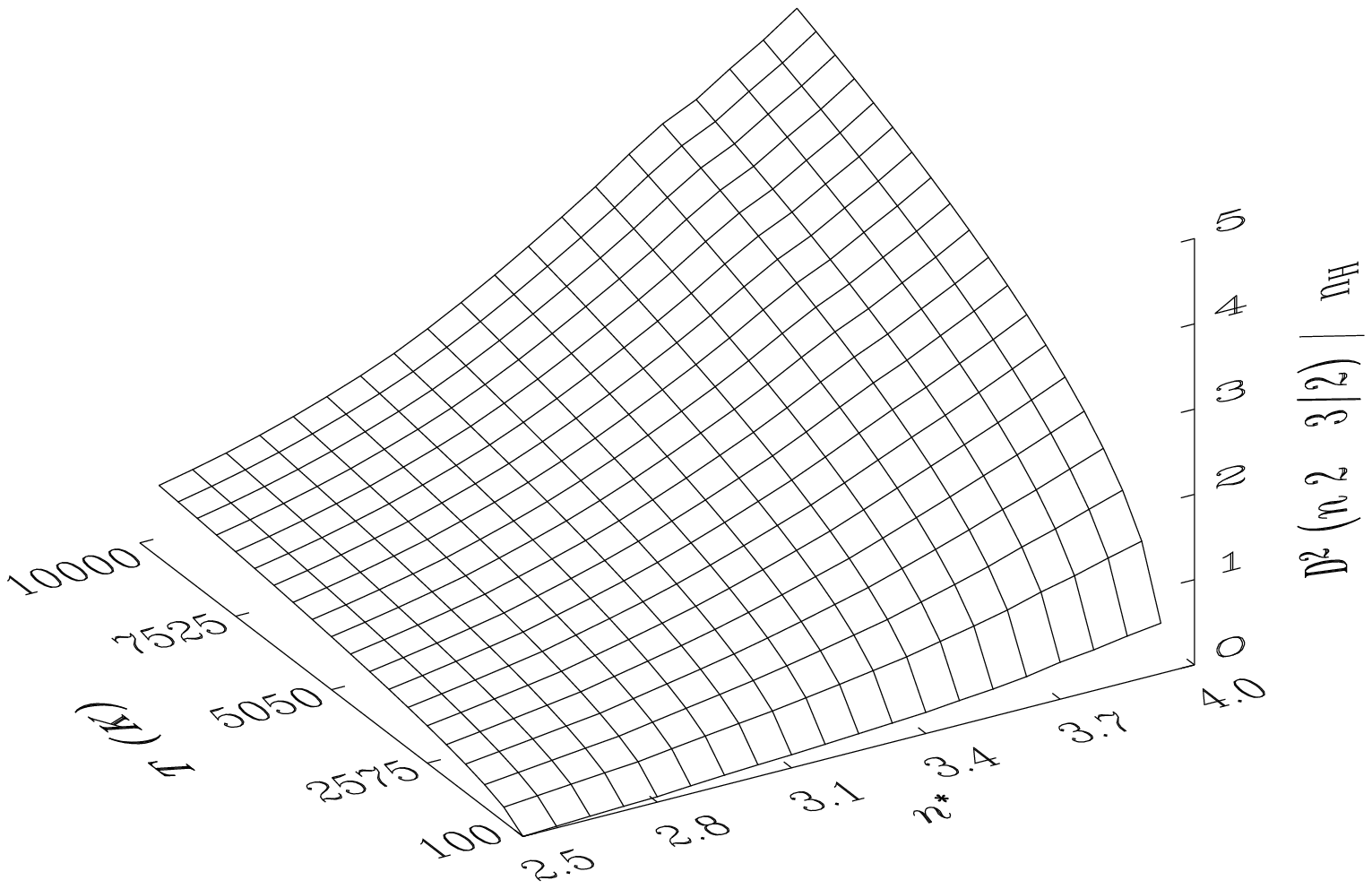}
\includegraphics[width=8 cm]{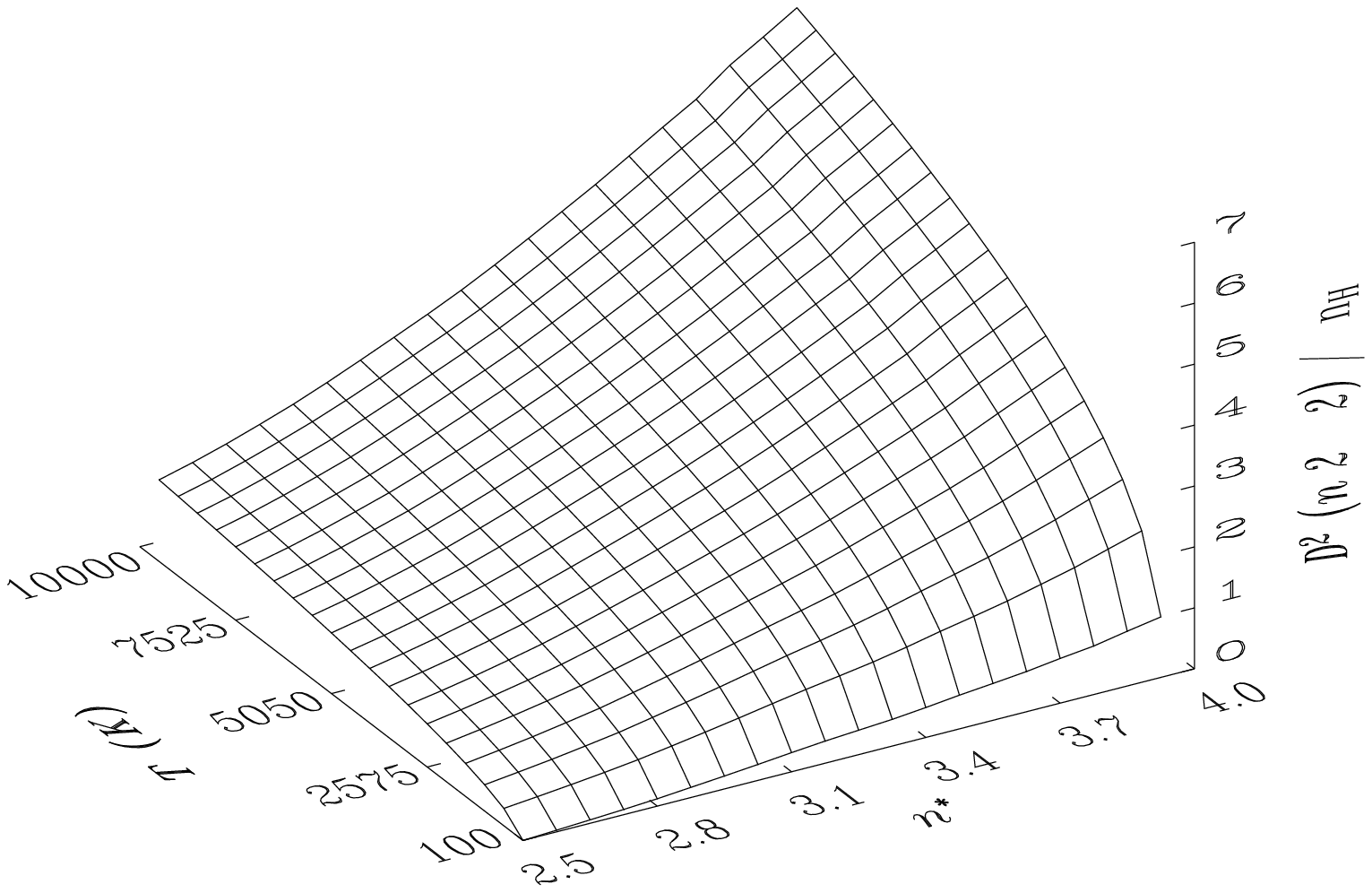}
\includegraphics[width=8 cm]{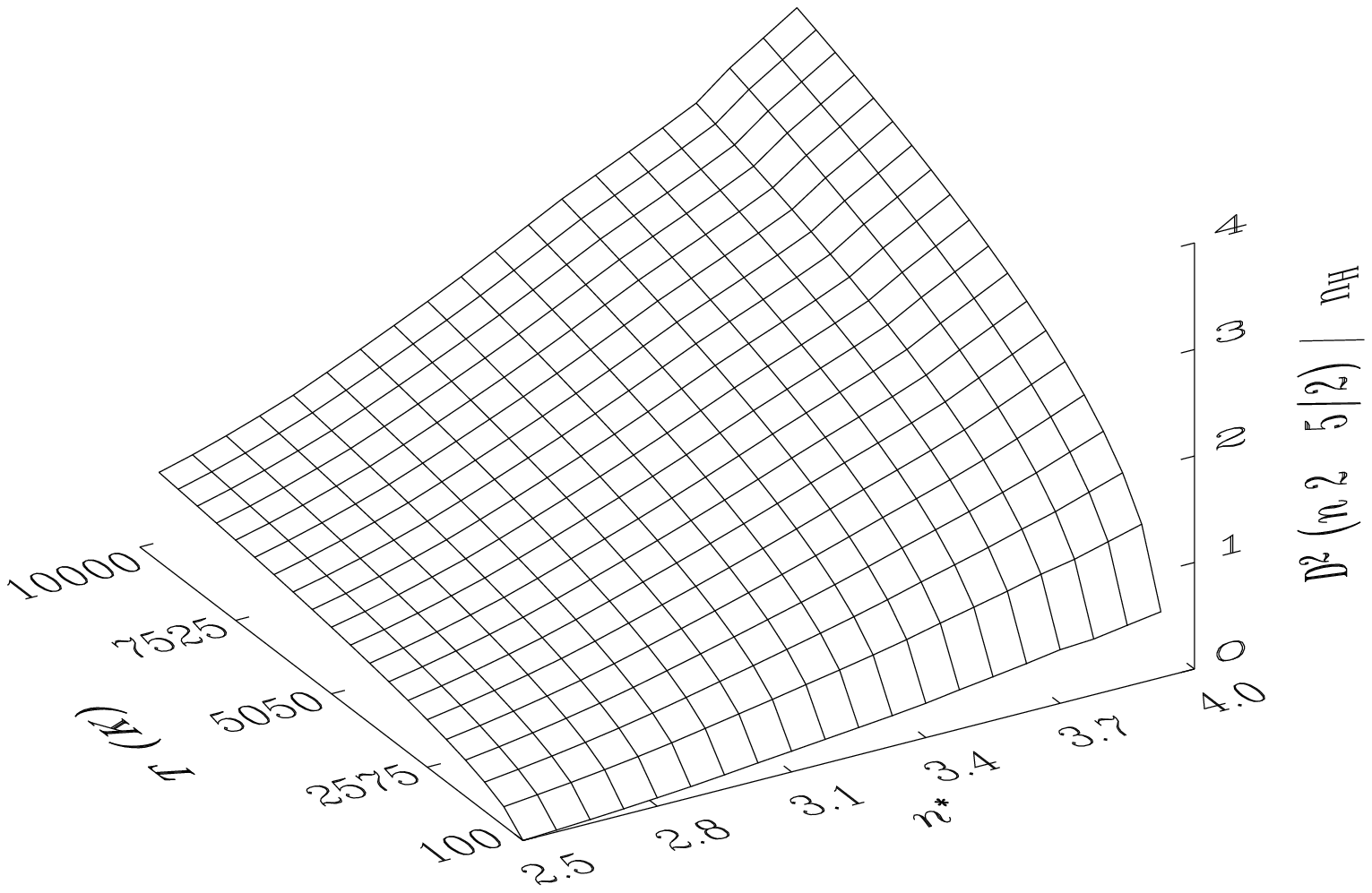}
\end{center}
\caption{Depolarization rates ($k$=2), per unit H-atom density, as  functions of temperature $T$ and $n^*$. For $l=2$, each figure:  $S=\frac{1}{2}$ and $J=\frac{3}{2}$; $S=0$ and $J=2$; $S=\frac{1}{2}$ and $J=\frac{5}{2}$. Depolarization rates are given in $10^{-14}$ $\textrm{rad.} \;  \textrm{m}^3 \; \textrm{s}^{-1}$.}
\label{depolarizationratefunction}
\end{figure}

\begin{figure}[htbp]
\begin{center}
\includegraphics[width=8 cm]{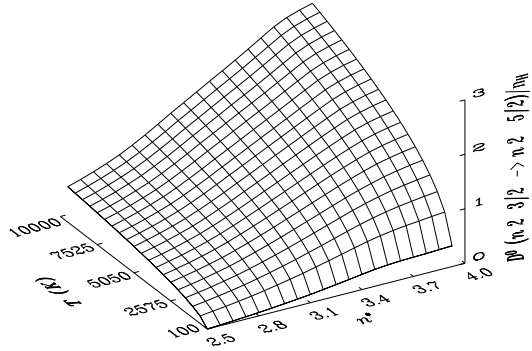}
\end{center}
\caption{Population transfer rate ($k$=0), per unit H-atom density, as a function of temperature $T$ and $n^*$. $l=2$, $S=\frac{1}{2}$, $J=\frac{3}{2}$ and $J'=\frac{5}{2}$.  Population transfer rate is given in $10^{-14}$ $\textrm{rad.} \;  \textrm{m}^3 \; \textrm{s}^{-1}$.}
\label{transferfunction0}
\end{figure}

\begin{figure}[htbp]
\begin{center}
\includegraphics[width=8 cm]{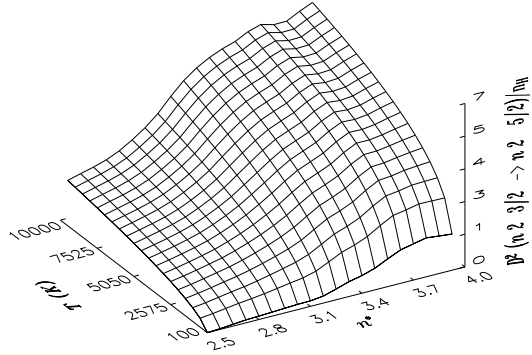}
\end{center}
\caption{Linear polarization transfer rate ($k$=2), per unit H-atom density,  as a function of temperature $T$ and $n^*$.  $l=2$, $S=\frac{1}{2}$, $J=\frac{3}{2}$ and $J'=\frac{5}{2}$. Linear polarization transfer rate is given in $10^{-15}$ $\textrm{rad.} \;  \textrm{m}^3 \; \textrm{s}^{-1}$.}
\label{transferfunction2}
\end{figure}

Figure \ref{depolarizationratefunction} shows the alignment depolarization rates ($k=2$) as a function of 
the local temperature $T$ and $n^*$ for $l=2$. The population transfer 
rates ($k=0$) and the linear polarization transfer rates ($k=2$) as a 
function of $T$ and $n^*$ are displayed in Figs \ref{transferfunction0} and \ref{transferfunction2}. All these rates increase with the temperature.  
For a temperature $\displaystyle T 
\leq $ 10000 K,  the destruction rate of alignment  $\displaystyle D^2(n \; 2
 \; 3/2)/n_H \leq 6  \times 10^{-14}$ $\textrm{rad.} \;  \textrm{m}^3 \; \textrm{s}
^{-1}$, $\displaystyle D^2(n \; 2 \; 2)/n_H \leq 8 \times 10^{-14}$ $\textrm{rad.}
 \;  \textrm{m}^3 \; \textrm{s}^{-1}$  and $\displaystyle D^2(n \; 2 \; 5/2)/ 
n_H \leq 5  \times 10^{-14}$ $\textrm{rad.} \;  \textrm{m}^3 \; \textrm{s}^{-1}$. 
The population transfer rate $\displaystyle D^0(n \; 2 \; 3/2  \to n \; 2 \; 
5/2)/n_H \leq 4  \times 10^{-14}$ $\textrm{rad.} \;  \textrm{m}^3 \; \textrm{s}^
{-1}$ and the linear  polarization transfer rate $\displaystyle D^2(n \; 2 \; 
3/2  \to n \; 2 \; 5/2)/n_H \leq 8  \times 10^{-15}$ $\textrm{rad.} \;  \textrm{m}^
3 \; \textrm{s}^{-1}$. These numerical values are given for  $\displaystyle 
n^* \leq 4$ which include most of the lines of  interest for the  second solar spectrum studies. The linear polarization transfer rates are smaller than the other 
rates.  In fact  equation (\ref{eq3})  shows that  $D^2(n \; 2 \; 
3/2  \to n \; 2 \; 5/2)$ is a linear combination of  $\zeta (n l J M_J \to n l J' M'_J, T)$. The coefficients of this linear combination have the sign of $(3M^2_J - J(J+1)) \times (3M'^2_J - J'(J'+1))$. Therefore these coefficients are sometimes positive and sometimes negative. Consequently, due to the  compensation between the different collisional rates  $\zeta (n l J M_J \to n l J' M'_J, T)$, $D^2(n \; 2 \; 3/2  \to n \; 2 \; 5/2)$ is  small compared with the other rates. For  similar reasons  circular polarization transfer rates ($k=1$) are negative for ($J=1/2$ $\to$ $J'=3/2$) and ($J=3/2$ $\to$ $J'=5/2$). This remark is in agreement with the negative  quantum chemistry orientation transfer rate  obtained by Kerkeni (\cite{Kerkeni2}) for Na D lines ($J=1/2$ $\to$ $J'=3/2$). We do not give our results concerning the circular polarization transfer because they are not needed for interpretation of the second solar spectrum . 

\section{Discussion}
The interactions of importance for the depolarization rates, \textbf{and the polarization and population transfer rates} 
 calculations are the intermediate-range interactions. In the corresponding regions accurate interaction potentials are required. Examination of
the RSU potential, used in this work, shows that long-range regions have the usual 
$R^{-6}$ behaviour but intermediate-regions have $ \sim R^{-10}$ behaviour. 
The Van der Waals potential is proportional to $R^{-6}$ at all separations and 
so that gives 
a good description of the problem only at long-range separations. The  Van der Waals 
potential underestimates the magnitude of the intermediate-range interactions. 
For this reason it can be seen why the calculations using the Van der Waals 
potential underestimate the depolarization and collisional transfer rates  
values. The results of Paper I showed that the RSU potential gave depolarization rates   which were in agreement ($<$ 20 \% for solar temperatures) with 
the quantum chemistry calculations.

Unfortunately,  there is neither  experimental nor quantum chemistry 
depolarization  and collisional transfer rates for $d$ states (at least to the authors' knowledge) to compare with. 
The main differences 
between the  RSU potentials and those from quantum chemistry, which are
more accurate, occur at  short-range interactions. We  verified that, as  for the $p$ states calculations in Paper I, these close collisions do not 
influence the computed depolarization  and collisional transfer rates for the $d$ states. We expect that a rather good agreement  with a full quantum mechanical treatment (difference less than 20 \%) would also occur for our present $d$ states results. 
\section{Conclusion}
The problem is to determine a great number of depolarization and collisional
transfer rates by isotropic collisions with H-atoms.  Our method was presented, tested and used with success \textbf{for} $p$ states in Paper I. In the present paper we have given  depolarization and collisional transfer rates  corresponding to $d$ states.  These results are needed to model the formation of observed lines, and thus interpret the observations in terms of the solar magnetic field.  The need is particularly strong for data for heavy complex atoms which are inaccessible to the quantum chemistry approach.  An extension to  $f$ atomic states ($ l = 3$) is a further interesting step in view of an extrapolation for $ l > 3$ states. Such an extension to higher $l$-values would be  useful for a global interpretation of the ``second solar spectrum''. This work is in progress. Adaptation and application of our theory to the determination of the depolarization and collisional transfer rates of singly ionized atoms lines by collisions with H-atoms will also be the subject of further papers.



\begin{thebibliography}{}
\bibitem[1991]{Anstee1}Anstee S.D., \& O'Mara B.J., 1991, MNRAS, 253, 549
\bibitem[1992]{Anstee2}Anstee S.D. PhD thesis, Univ. Queensland, 1992
\bibitem[1995]{Anstee3}Anstee S.D., \& O'Mara B.J., 1995, MNRAS, 276, 859
\bibitem[1997]{Anstee4}Anstee S.D., O'Mara B.J., \& Ross J.E., 1997, MNRAS, 284, 202
\bibitem[1997]{Barklem1}Barklem P.S., \& O'Mara B.J., 1997, MNRAS, 290, 102
\bibitem[1998]{Barklem2}Barklem P.S., \& O'Mara B.J., 1998, MNRAS, 296, 1057
\bibitem[1998]{Barklem3}Barklem P.S. PhD thesis , Univ. Queensland, 1998
\bibitem[2000]{Barklem4}Barklem P.S., Piskunov N., \& O'Mara B.J., 2000, A\&A, 142, 467
\bibitem[2002]{Bommier3}Bommier V., \& Molodij G., 2002, A\&A, 381, 241 
\bibitem[2003]{derouich1}Derouich M., Sahal-Bréchot S., Barklem P.S., \& O'Mara B.J., 2003, A\&A, 404, 763
\bibitem[2002]{Kerkeni2}Kerkeni B., 2002, A\&A, 390, 791 
\bibitem[2002]{manso}Manso Sainz R., \& Landi Degl'Innocenti E., 2002, A\&A, 394, 1093
\bibitem[1961]{Messiah}Messiah A., 1961, Mécanique Quantique (Paris: Dunod)
\bibitem[1974]{roueff}Roueff E., 1974, J.\ Phys.\ B, 7, 185 
\bibitem[1977]{Sahal2}Sahal-Bréchot S., 1977, ApJ., 213, 887
\bibitem[1997]{Stenflo}Stenflo J.O., \& Keller C.U., 1997, A\&A, 321, 927
\end{thebibliography}
\end{document}